\documentclass[aps,prl,a4paper,10pt,twocolumn,showpacs,longbibliography,floatfix,superscriptaddress,amsmath,amsfonts,amssymb,nofootinbib]{revtex4-2}
\usepackage{graphicx}
\usepackage{calc}
\usepackage{array}
\usepackage{amsmath, amssymb}
\usepackage{natbib}
\usepackage{hyperref}
\usepackage{overpic}
\usepackage{comment}
\usepackage[dvipsnames]{xcolor}

\DeclareMathOperator{\dd}{\mathrm{d}\!}

\DeclareMathOperator{\Tr}{\mathrm{Tr}}

\newenvironment{acknowledgment}
{\vspace{0.5cm}\noindent\textbf{Acknowledgments ---}}
{\par\vspace{0.5cm}}

\begin{document}

\title{Transverse Magnetic Response from Orbitally Polarized Cooper Pairs in Elemental Superconductors}

\author{G\'abor Csire}
%\email{csire.gab@gmail.com}
\affiliation{CNR-SPIN, c/o Università di Salerno, IT-84084 Fisciano (SA), Italy}

\author{Maria Teresa Mercaldo}
\affiliation{Dipartimento di Fisica “E. R. Caianiello”, Università di Salerno, IT-84084 Fisciano (SA), Italy}

\author{Bal\'azs \'Ujfalussy}
\affiliation{Wigner Research Centre for Physics, Institute for Solid State Physics and Optics, H-1525 Budapest, Hungary}

\author{Carmine Ortix}
\affiliation{Dipartimento di Fisica “E. R. Caianiello”, Università di Salerno, IT-84084 Fisciano (SA), Italy}
\affiliation{CNR-SPIN, c/o Università di Salerno, IT-84084 Fisciano (SA), Italy}

\author{Mario Cuoco}
\affiliation{CNR-SPIN, c/o Università di Salerno, IT-84084 Fisciano (SA), Italy}

\date{\today}
\begin{abstract}
We demonstrate how crystalline symmetry lowering, as for instance through strain, allows elemental superconductors such as vanadium and niobium to realize spin-singlet orbitally polarized Cooper pairs composed of electrons with identical orbital moments.
Using superconducting density functional theory, we show that lowering of trigonal symmetry to $\mathcal{C}_s$, thus keeping only a single mirror plane, activates interorbital pairing in bulk and (111) surfaces, with a pronounced surface enhancement. In a magnetic field, the resulting orbitally polarized superconducting state leads to a novel transverse magnetic response. 
For in--plane field orientations that break the remaining mirror symmetry, a sizable orbital magnetization emerges perpendicular to the applied field. We show that this effect is a direct consequence of equal--orbital-moment Cooper pairing, providing an experimentally accessible signature of this state. Our results establish strained elemental superconductors as a minimal material platform for superconducting orbitronics.
\end{abstract}

\maketitle

\section{Introduction}

\begin{figure*}
    \centering
    \includegraphics[width=1\textwidth]{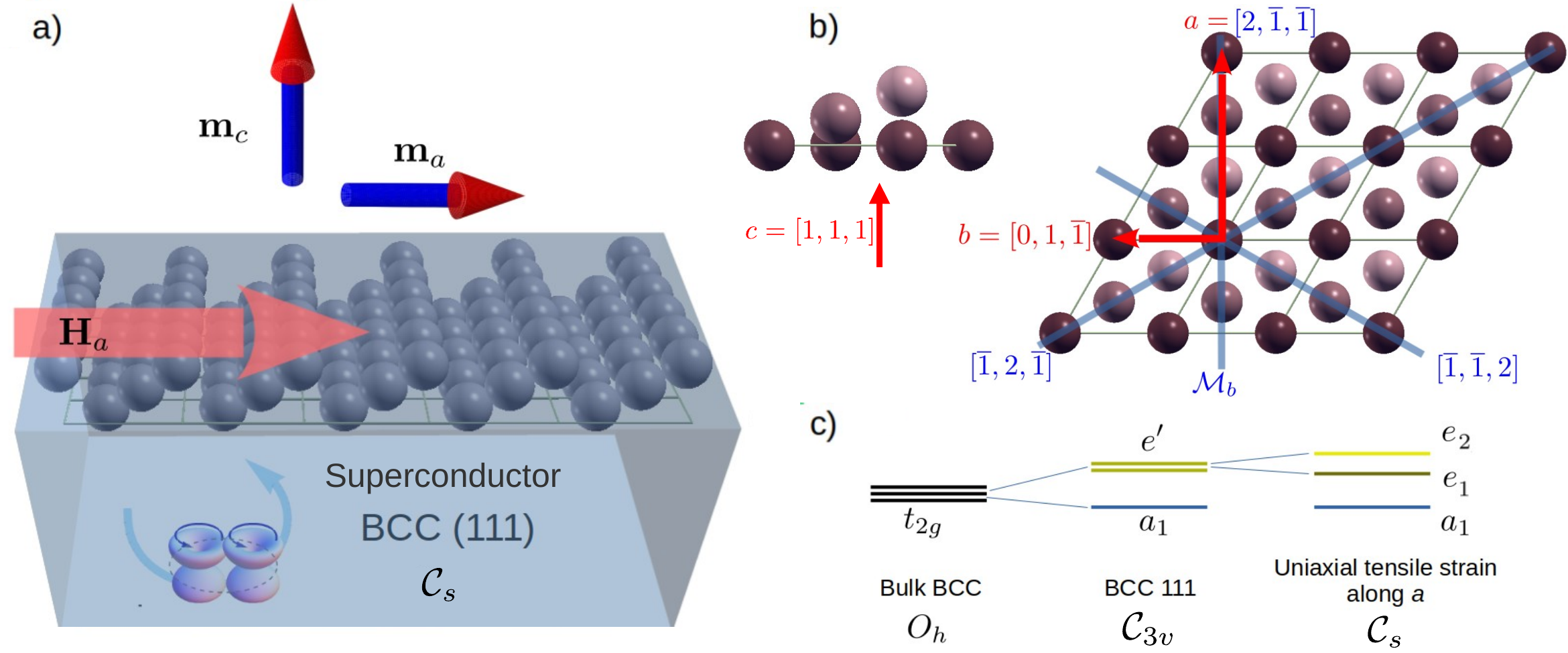}
    \caption{\textbf{Strain-induced transverse orbital magnetic response in elemental superconductors.}
    (a) Schematic showing that an in-plane magnetic field induces an out-of-plane orbital moment on the strained [111] surface. The application of the strain field can lower the trigonal symmetry to $\mathcal{C}_s$ symmetry thus keeping only a single mirror plane at the surface. 
    (b) Unit cell of bulk bcc crystal stacked along [111] (left) and top view showing crystallographic axes $\hat a$, $\hat b$ and the $\mathcal M_b$ mirror plane, the only one remaining after strain applied along $\hat a$. 
    (c) Evolution of orbital states at the $\Gamma$ point from bulk $t_{2g}$ to trigonal ($\mathcal{C}_{3v}$) surface states, and their further splitting under strain to $\mathcal{C}_s$, where we retain the label $e$ to indicate their common origin from the $\mathcal{C}_{3v}$ doublet.
    }
    \label{fig:schematic}
\end{figure*}

Hybrid heterostructures of conventional superconductors and magnetic materials can generate spin-polarized Cooper pairs for superconducting spintronics~\cite{Linder2015, Amundsen2024}, yet device development remains challenging due to the inherent complexity of integrating magnetism with superconductivity. 
However, breaking crystalline symmetries—such as mirror or rotational symmetries—provides an alternative route to generating Cooper pairs with finite angular momentum, even in systems with negligible spin-orbit coupling \cite{Mercaldo2023}.
In this scenario, electrons can form orbitally polarized Cooper pairs, pairing with equal atomic orbital moment while remaining in a spin-singlet configuration.
Unlike in superconducting spintronics, this angular momentum arises from the orbital rather than the spin degrees of freedom, reflecting a mechanism based on orbital structure instead of spin polarization.
This perspective underpins the concept of superconducting orbitronics\cite{Mercaldo2023}, where orbital supercurrents -- flows of Cooper pairs carrying finite orbital angular momentum~(OAM) -- can be generated and manipulated, offering a promising platform for novel types of superconducting electronics and quantum information processing.

Orbital degrees of freedom become particularly relevant in systems lacking inversion symmetry,
where the orbital Rashba coupling~\cite{Park2011, Park2012} ties the atomic OAM to the crystal momentum. 
In multiorbital materials, this coupling is electrically tunable and can lead to profound consequences.
It can drive a 0–\(\pi\) orbital transition, where the superconducting order parameter 
acquires opposite signs on different orbitals~\cite{Mercaldo2020}, producing 
measurable signatures in tunneling spectroscopy~\cite{Mercaldo2021}, magnetic 
response~\cite{Bours2020}, and Josephson transport~\cite{Guarcello2022}. 
Moreover, even conventional $s$-wave spin-singlet superconductors with low 
crystalline symmetry may host vortices carrying orbitally polarized supercurrents without any applied magnetic field~\cite{Vortex2022}. 
The orbital Rashba coupling  can also induce an orbital Edelstein effect~\cite{Chirolli2022, Ando2024, saunderson2025},
yielding an orbital polarization that surpasses its spin counterpart by an order of magnitude. 
In acentric multiorbital superconductors, interorbital spin-triplet pairing can stabilize
an even-parity s-wave topological phase~\cite{Fukaya2018}, supporting 0-, \(\pi\)-, and \(\varphi\)-Josephson couplings
with enhanced higher harmonics~\cite{Fukaya2020, Fukaya2022},
and can also drive distinctive spin- and orbital-texture transitions across the superconducting state~\cite{Fukaya2019}.
Intriguingly, in several acentric or non-symmorphic s-wave superconductors, the onset of time-reversal symmetry breaking can be accounted within a framework where interorbital pairing provides the conditions and mechanisms for its emergence~\cite{Csire2018, Ghosh2020, Csire2022, Ramires2022}.
Moreover, the relevance of orbital selective or interorbital pairing was already identified in many unconventional superconductors including the nickelates~\cite{Jakub2025}, kagome superconductors~\cite{Huang2025}, iron-selenide based superconductors~\cite{Nicaetal2017},
cuprates\cite{saunderson2025c} and Sr$_2$RuO$_4$~\cite{Ramires2019, CsireFukaya2026}.

Despite the large amount of work on multiorbital superconductors and the potential of achieving orbitally polarized Cooper pairs, their direct experimental evidence remains elusive. In particular, the identification of suitable material platforms and clear-cut experimental signatures are an open challenge. Here, we address this problem by predicting a previously unexplored magnetic response that provides a direct and experimentally accessible route to detecting such type of Cooper pairs. We show that this effect uniquely arises from orbitally polarized Cooper pairs, thereby offering a clearcut probe of their existence. Moreover, they can emerge in simple elemental superconductors, bringing superconducting orbitronics within reach of existing materials and techniques.

To this aim, in this work we study elemental superconductors, such as V and Nb, and their 111
surfaces using superconducting density functional theory~\cite{Oliveira1988, Csire2015, Csire2018kkr}. We show that high orbital-moment
Cooper pairs can exist provided the crystalline symmetry is appropriately reduced.
Importantly, we take significant steps toward experimental detection by considering
the effect of an external field, accounting for both spin and orbital coupling. We
quantitatively predict a novel effect tied to the presence of high orbital-moment
Cooper pairs.
Specifically, when an in-plane magnetic field is applied to the strained material with only a single remaining mirror symmetry, the induced magnetic moment acquires a sizable out-of-plane component, as schematically illustrated in Fig.~\ref{fig:schematic}a. 

\section{Interorbital pairing in trigonal systems with broken crystalline symmetry}

Vanadium and niobium crystallize in the body-centered cubic (bcc) structure, and their electronic states near the Fermi level are primarily derived from the $t_{2g}$ orbitals, resulting in similar Fermi surfaces. 
The lattice constants of vanadium and niobium are $a = 3.03~\text{\AA}$ and $a = 3.30~\text{\AA}$, respectively. 
Their superconducting gaps are approximately $0.8~\text{meV}$ for V and $1.5~\text{meV}$ for Nb.
We shall study bcc layers stacked along the [111] direction, where the stacking sequence consists of triangular planes with ABC-type stacking as shown in Fig.~\ref{fig:schematic}b.
To describe the orbital character relative to this stacking, we define an orthonormal trigonal reference frame adapted to this direction, with $\hat{c}$ along [111], $\hat{a}$ in-plane [2$\bar{1} \bar{1}$] along the strain direction, and $\hat{b}$ in-plane perpendicular to $\hat{a}$ as [01$\bar{1}$].
In this frame, the cubic $t_{2g}$ orbitals of the bulk can be expressed as a singlet along the trigonal axis and a doublet in the plane perpendicular to it. The orbital along the trigonal axis is
\begin{equation}
a_1 = \frac{1}{\sqrt{3}} \left( d_{xy} + d_{yz} + d_{zx} \right),
\end{equation}
while the two in-plane orbitals forming a doublet are
\begin{align}
e_1 &= \frac{1}{\sqrt{6}} \left( 2 d_{xy} - d_{yz} - d_{zx} \right),\\
e_2 &= \frac{1}{\sqrt{2}} \left( d_{yz} - d_{zx} \right).
\end{align}
The doublet is degenerate in the unstrained crystal, and any in-plane strain along $\hat a$ lifts this degeneracy (Fig.~\ref{fig:schematic}c).
The same reasoning applies to surfaces derived from these layers.
These orbitals form an $L = 1$ multiplet manifold in the trigonal crystalline environment.

In order to perform realistic, material-specific calculations, we solve the density-functional (i.e., Kohn--Sham (KS)) Dirac--Bogoliubov--de~Gennes (DBdG) Hamiltonian\cite{Capelle1999} using band-structure methods\cite{Csire2015, Csire2018kkr} that yield the full Green's function, as described in the Methods section.
The key advantage of this Green's-function approach is its ability to incorporate a realistic Fermi surface, orbital composition, and spin--orbit coupling together with an arbitrary superconducting pairing model, while simultaneously treating semi-infinite geometries appropriate for modeling surfaces.
This framework naturally enables accurate predictions of pairing amplitudes in an orbital basis, 
$\chi_{\alpha\beta}^{\mu\nu} = \langle c_{\mu\alpha}(\mathbf{k}) c_{\nu\beta}(-\mathbf{k}) \rangle$,
where $\mu,\nu$ label the orbitals and $\alpha,\beta$ denote spin indices.

The equal orbital moment operator is most naturally formulated by focusing directly on the $J=2$ (quintet) sector, which provides the only nontrivial spin--singlet pairing channel with finite orbital angular momentum. For the examined multiorbital system the $L=1$ manifold is described by Wannier orbitals ($a_1$,$e_1$,$e_2$). Then, by suitably combining these configurations one can construct  $L=1$ orbitals with nonvanishing orbital projection along the trigonal axes. Their combination leads to states with $J=0,1,2$ and we are interested on the configurations with total orbital angular momentum $J=2$. The quintet states, e.g. $|2,J_i\rangle$ ($i=\hat a, \hat b, \hat c$), are then obtained by combining orbital degrees of freedom using Clebsch–Gordan coefficients ~\cite{Mercaldo2023}. This yields specific linear combinations of intra--orbital and inter-orbital pairing operators. 
For convenience and clarity, we focus only on the relevant interorbital components of the maximal projection operators for the quintet configurations along the trigonal axes $\hat a, \hat b, \hat c$ as they provide the necessary phase relationships to establish a net orbital moment of the Cooper pairs. The resulting operators are expressed in the following form: 
\begin{align}
\mathcal{O}_{a}(\mathbf{k}) &= c_{e_2\uparrow}(\mathbf{k})\, c_{a_1\downarrow}(-\mathbf{k}) + c_{a_1\uparrow}(\mathbf{k})\, c_{e_2\downarrow}(-\mathbf{k}), \\[0.5em]
\mathcal{O}_{b}(\mathbf{k}) &= c_{a_1\uparrow}(\mathbf{k})\, c_{e_1\downarrow}(-\mathbf{k}) + c_{e_1\uparrow}(\mathbf{k})\, c_{a_1\downarrow}(-\mathbf{k}), \\[0.5em]
\mathcal{O}_{c}(\mathbf{k}) &= c_{e_1\uparrow}(\mathbf{k})\, c_{e_2\downarrow}(-\mathbf{k}) + c_{e_2\uparrow}(\mathbf{k})\, c_{e_1\downarrow}(-\mathbf{k}).
\end{align}
When considering the pairing amplitude in momentum space, nonzero interorbital contributions do exist, however, after integration over the Brillouin zone, the net interorbital pair correlations vanish due to symmetry constraints.
To generate a finite net interorbital pairing, the crystal symmetry must be lowered to $\mathcal{C}_s$, which we achieve by applying uniaxial tensile strain. 
When mirror symmetries are broken except the $\mathcal M_b$ mirror plane, the symmetric orbital-mixing term $\mathcal{O}_{b}$ is allowed by symmetry.
It has the same orbital-coupling pattern as the OAM operator $L_b$: it connects exactly the same pairs of orbitals, but in a symmetric (rather than antisymmetric) manner.
For convenience and clarity of presentation, the interorbital components of the maximal projections will be normalized with the sum of intraorbital singlet pair correlations.

\begin{figure}[!htb]
    \centering
    \includegraphics[width=1\columnwidth]{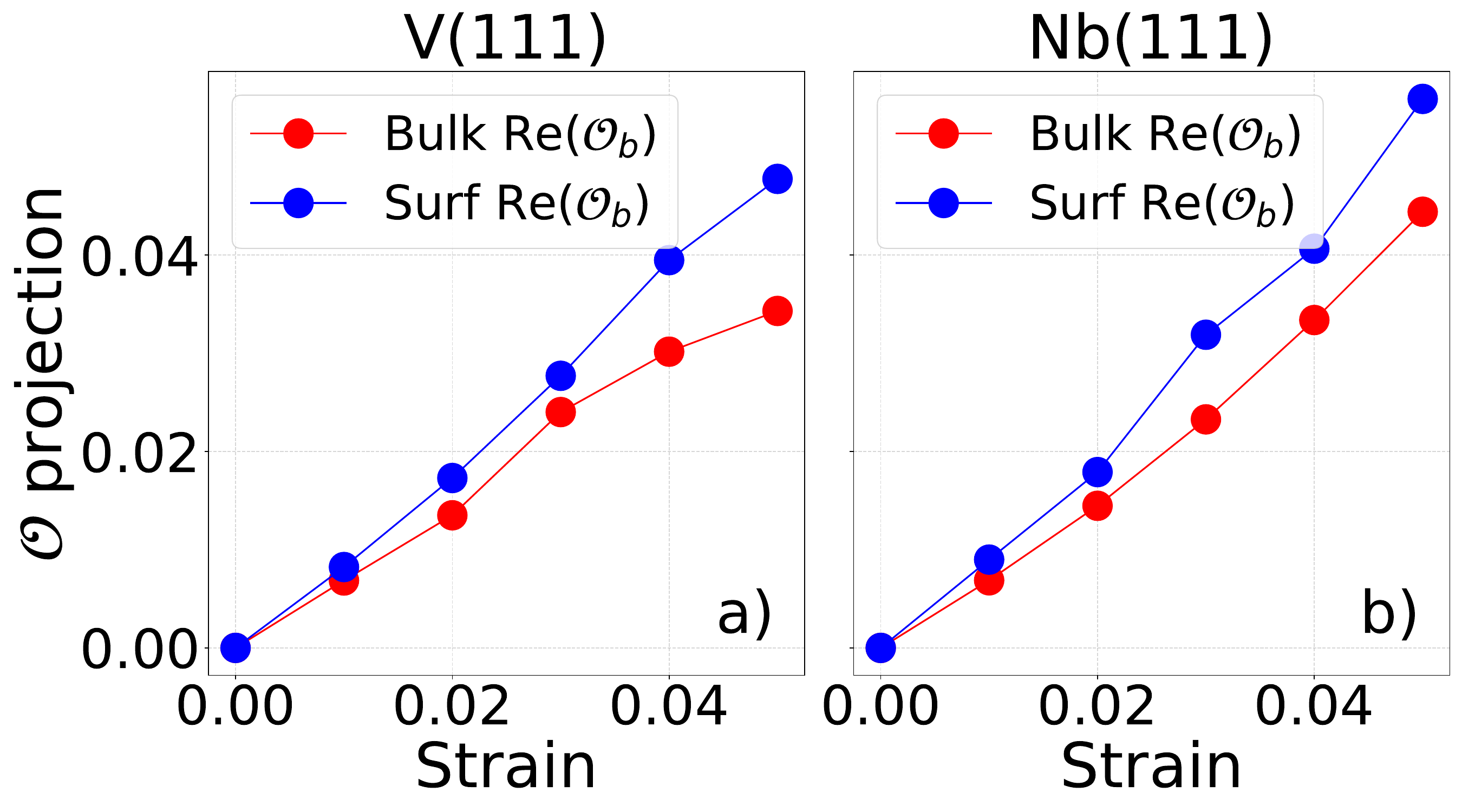}
    \caption{\textbf{Strain-driven interorbital pairing in elemental superconductors.}
    Interorbital components of the maximal projections for (a) V(111) and (b) Nb(111) as a function of tensile strain applied along the $\hat{b}$ axis. Results are shown for both bulk and surface cases. Only non-zero projection elements are displayed. Strain is expressed as the relative percentage change of the unit cell’s long diagonal.}
    \label{fig:proj}
\end{figure}

To examine the presence of high orbital-moment Cooper pairs, we carried out KSDBdG calculations for the bulk (111 layers) and the (111) surface under different strain conditions. The strain is applied as a uniaxial stretch along the $a$ axis, preserving the $\mathcal M_b$ mirror symmetry.
The resulting non-zero orbital projections are shown in Fig.~\ref{fig:proj} for vanadium (a) and niobium (b).
Interorbital pairing emerges as the applied strain removes the remaining mirror symmetries. However, because the $\mathcal M_b$ mirror symmetry is preserved, this pairing manifests specifically through a finite, real $\mathcal O_b$ projection, which preserves time-reversal symmetry.
Comparing the bulk and surface results, we find that for both V and Nb the $\mathcal O_b$ projection is significantly enhanced at the surface, consistent with the additional symmetry breaking associated with inversion symmetry loss.
By comparing vanadium and niobium, one can also observe larger $\mathcal O_b$ values in Nb, suggesting a role of atomic spin–orbit coupling (SOC).
Being a heavier element, Nb naturally has stronger SOC than V, which is still relatively weak on the usual electronic structure scale (eV) but it can have a noticeable effect on the superconducting energy scale (meV).
Indeed, as shown in the Supplemental Material, scaling down the SOC strength in Nb leads to a reduction of the $\mathcal O_b$ projection.
This demonstrates that while SOC is not a necessary ingredient for interorbital pairing, it can enhance its magnitude.

\section{Transverse magnetic response from orbitally polarized Cooper pairs}

\begin{figure}[!htb]
    \centering
    \includegraphics[width=1\columnwidth]{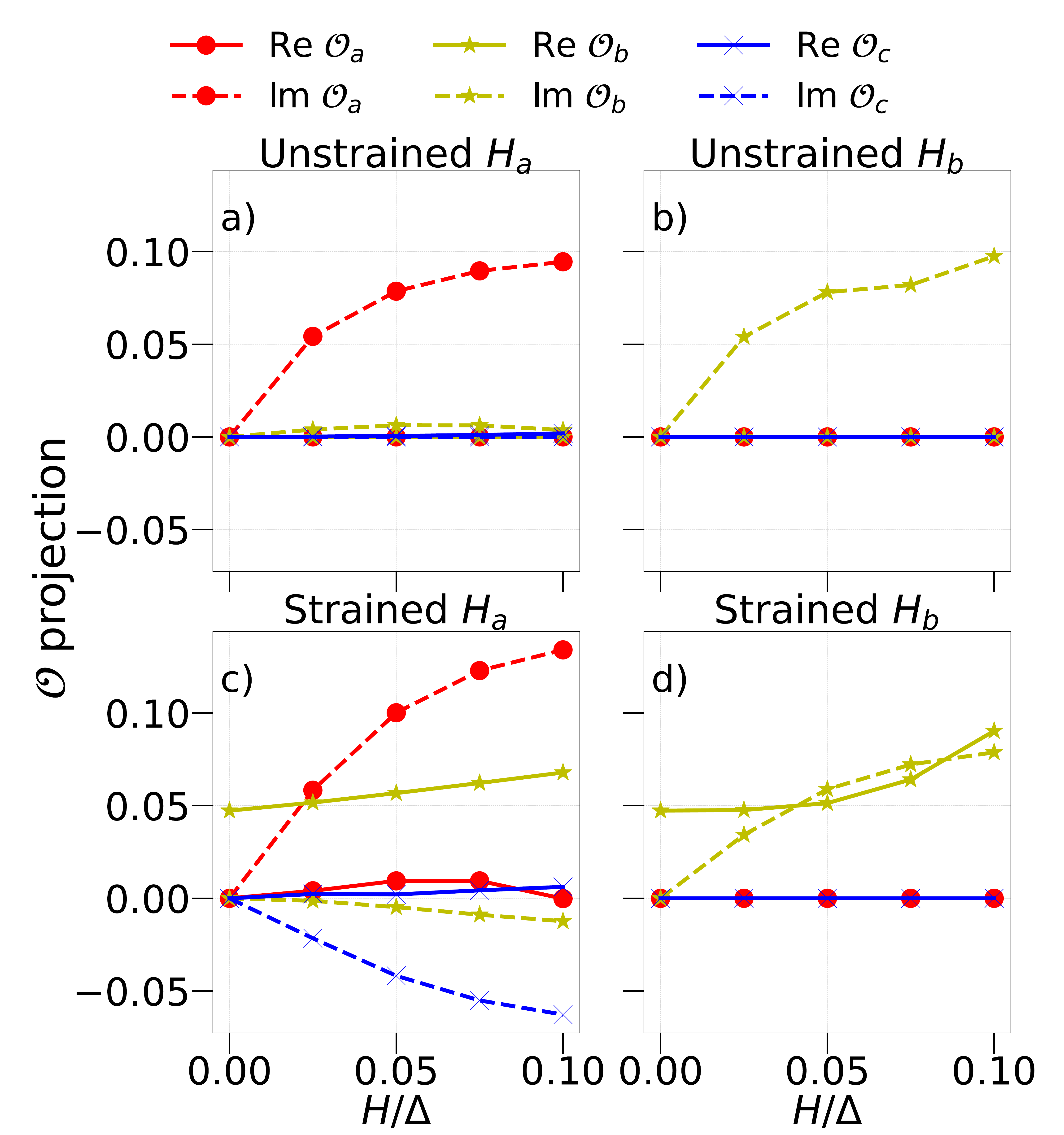}
    \caption{\textbf{Interorbital components of the maximal projections for V(111) surface as a function of external magnetic field.} The field was applied in two in-plane orientations on the surface, along $a$ and $b$. Panels (a--b) correspond to the unstrained case, and panels (c--d) correspond to the strained case, each with different in-plane orientations of the external field.}
    \label{fig:VprojH}
\end{figure}

\begin{figure*}%[!htb]
    \centering
    \includegraphics[width=0.9\textwidth]{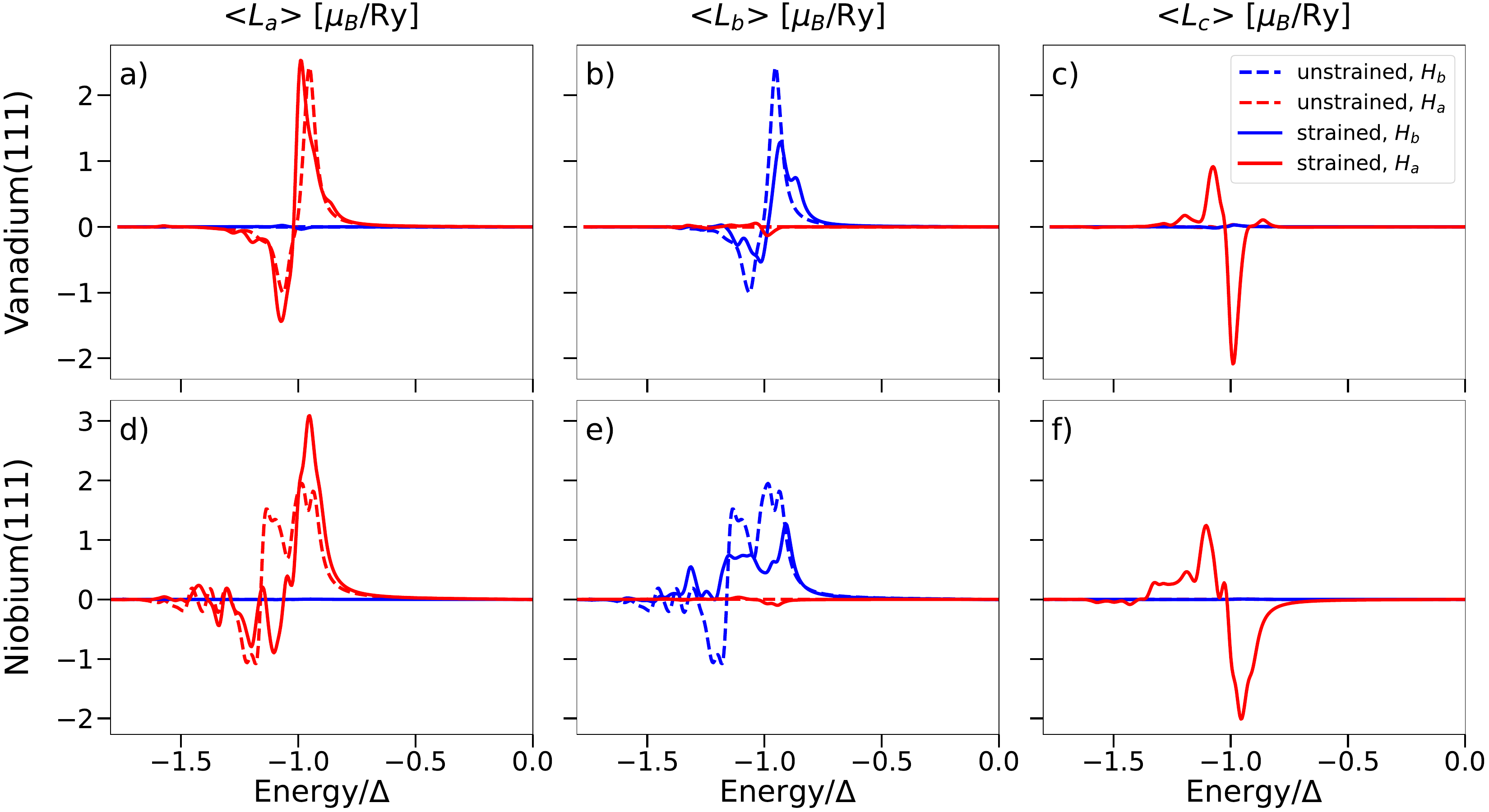}
    \caption{\textbf{Energy-resolved orbital moment along different directions under external magnetic fields within the surface plane.} $H = 0.1\Delta$ was applied in two in-plane orientations, along $a$ and $b$. Results are shown for unstrained (C$_{3v}$, dashed line) and strained (C$_S$, solid line) cases. Panels (a--c) correspond to the V 111 surface, and panels (d--f) to the Nb 111 surface. Red curves indicate that the applied external field lies within the mirror plane (preserved under strain), whereas blue curves correspond to a field applied perpendicular to the mirror plane.}
    \label{fig:orbmom}
\end{figure*}

We now turn to studying the effect of an external magnetic field, focusing on vanadium as a candidate material.
We emphasize that the external magnetic field affects both the spin and orbital degrees of freedom.
Consequently, we account for both the spin and orbital Zeeman effects.
As we will show, including the orbital Zeeman term gives rise to qualitatively novel physical phenomena.
Since interorbital pairing is more pronounced at the strained surface, we perform our analysis on the (111) surface using a semi-infinite geometry.
The magnetic field is applied in-plane along two specific directions: one within the preserved $\mathcal M_b$ mirror plane and the other perpendicular to it.
The analysis is carried out for both the unstrained ($\mathcal{C}_{3v}$) and strained ($\mathcal{C}_s$) cases.
As before, we compute the interorbital components of the maximal orbital projections, and the results are presented as a function of the applied external field in Fig.~\ref{fig:VprojH}.
We underline that the external magnetic field interacts with both the spin and the orbital motion.
One can observe that the magnetic field generates the imaginary parts of $\boldsymbol{\mathcal{O}}$, while strain induced only real components.
This difference arises because the external magnetic field breaks time-reversal symmetry.

In the unstrained case (Fig.~\ref{fig:VprojH}a and b), no out-of-plane component of the $\boldsymbol{\mathcal{O}}$ projection is induced.
The $\boldsymbol{\mathcal{O}}$ projection essentially aligns with the direction of the applied field,
and increase monotonically with increasing field strength.
However, when the field is applied along the strain direction (the $a$ axis), all mirror symmetries are broken, which gives rise to a finite, albeit small, $\mathcal{O}_b$ projection.

A more interesting behavior emerges when the external magnetic field is applied within the strained surface, as shown in Fig.~\ref{fig:VprojH}c and d.
When the field is applied perpendicular to the $\mathcal M_b$ mirror plane, an imaginary $\mathcal O_b$ projection is induced along the same direction (Fig.~\ref{fig:VprojH}d), however, its magnitude is reduced compared to the unstrained case.
At the same time, the presence of strain generates a finite real component of $\mathcal O_b$.
A much richer response is observed when the magnetic field lies within the $\mathcal M_b$ mirror plane.
In this configuration, an imaginary $\mathcal O_a$ projection is induced along the field direction (Fig.~\ref{fig:VprojH}c), with a magnitude that is enhanced relative to the unstrained case.
In addition, strain again produces a finite real $\mathcal O_b$ component, and finite magnitudes for all components of $\boldsymbol{\mathcal O}$.
Most notably, a very significant out-of-plane imaginary component $\mathcal O_c$ is induced, with a sign opposite to that of the in-plane components.
Since this out-of-plane component $\mathcal O_c$ is mostly imaginary, it breaks time-reversal symmetry, therefore, it contributes to an OAM along the out-of-plane direction. 

Having observed that strain redistributes the OAM of Cooper pairs anisotropically under an external magnetic field, we now calculate the energy-resolved OAM for different field directions, as shown in Fig.~\ref{fig:orbmom}. The strained results presented in Fig.~\ref{fig:VprojH} and Fig.~\ref{fig:orbmom} were calculated for a 5\% uniaxial tensile strain applied along the $a$ axis.
For better comparison between vanadium (Fig.~\ref{fig:orbmom}a–c) and niobium (Fig.~\ref{fig:orbmom}d–f), the energy axis is rescaled by the respective superconducting gap magnitudes.
In the absence of strain, the OAM aligns with the direction of the applied magnetic field.
Once strain breaks the mirror symmetries, leaving only the 
$\mathcal M_b$ mirror, qualitatively new behavior emerges: a strong in-plane anisotropy of the OAM develops, and, most strikingly, an out-of-plane OAM appears only when the magnetic field lies within the $\mathcal M_b$ mirror plane, thereby breaking the remaining mirror symmetry in the strained system.
The most pronounced changes occur near the superconducting gap edge.
Moreover, the pairing correlations shown in Fig.~\ref{fig:VprojH} are in excellent qualitative agreement with the corresponding orbital-angular-momentum response.
Together, these observations provide strong evidence that the induced out-of-plane OAM generated by a transverse in-plane magnetic field originates from strain-driven, orbitally polarized Cooper pairs.

\section{Discussion and Conclusions}

Due to the breaking of mirror and rotational symmetries, Cooper pairs with an out-of-plane component of the orbital moment are induced, but its net value over the Brillouin zone vanishes in the absence of an external field because time-reversal symmetry is preserved.
When an in-plane magnetic field is present, a finite net out-of-plane magnetization emerges.
Crucially, this effect appears only in the superconducting state, providing a direct signature of strain-driven, orbitally polarized Cooper pairs.

Our findings establish a distinct physical scenario, as an induced magnetization transverse to an applied magnetic field is generally not expected in superconductors with high crystalline symmetry and predominantly single-orbital band structures. In such systems, rotational and mirror symmetries render the magnetic response tensor effectively diagonal, enforcing collinearity between the applied field and the induced magnetization.

Referring to the normal state properties, although a transverse response may arise in the presence of strong spin–orbit coupling, reduced crystalline symmetry, or due to magnetic exchange, these conditions are absent in the metallic phases of the elemental superconductors considered here. In particular, even under strain, our findings show that the normal state exhibits zero net induced magnetization, precluding any transverse response arising from the underlying metallic behavior and facilitating the detection of the effect.
In view of these aspects, our results identify a clear-cut effect uniquely associated with superconductors hosting high orbital-moment Cooper pairs.

Regarding possible experimental realizations with elemental superconductors, we note that Nb(111) can be grown successfully with a sufficiently precise control over surface quality \cite{Coupeau2015,Goedecke2023}. 
Instead, the realization of V(111) thin films though experimentally feasible is more challenging, as bcc vanadium naturally favors the (110) orientation. Early studies on growth over (111) substrates such as Cu, Ag, and Pt show that the (111) phase can be stabilized under controlled conditions.
More recent work on epitaxial heterostructures, such as Au/V systems, demonstrates that high-quality V(111) films can be achieved while preserving robust superconductivity with near-bulk $T_c$ \cite{Wei2016}. 

A direct application of our findings can be also found in the superconducting phase of the Pb/Si(111) system. In this case, the striped incommensurate phase breaks the higher symmetry of the ideal $\sqrt{3} \times \sqrt{3}$ reconstruction \cite{Menard2017}, reducing it to a lower point group such as $\mathcal{C}_s$. 
As predicted, this reduction in surface symmetry can naturally give rise to Cooper pairs with a finite orbital moment and to a transverse magnetic response under an in-plane applied magnetic field.

We studied bcc(111) elemental superconductors, but many fcc(111) materials, such as titanium and lead, are also superconducting.
Both bcc(111) and fcc(111) surfaces share triangular atomic arrangements, however, bcc(111) is more open, with lower surface coordination and closer interlayer spacing.
This weaker confinement along the surface normal may allow strain-induced symmetry breaking to generate a larger out-of-plane orbital moment,
while the denser fcc(111) surface more strongly confines orbitals in-plane, resulting in a smaller out-of-plane moment.
Therefore, the effect is likely stronger on bcc(111) surfaces. 

Other promising materials candidates for the suggested effect include 111-oriented oxide interfaces, such as SrTiO$_3$\cite{Udit2019}, KTaO$_3$\cite{Ren2022,Mallik2022}, and LaAlO-SrTiO-based heterostructures \cite{Monteiro2017,Davis2018}, which combine multiorbital superconductivity with low crystalline symmetry \cite{les23,Mercaldo2023A}.
Transition metal dichalcogenides can also host low-symmetry multiorbital electronic phases with only a single remaining mirror symmetry, most notably through the 1T $\rightarrow$ 1T$^{'}$ structural transition. 
Importantly, superconductivity can emerge in this phase, as observed in materials such as MoTe$_2$ and WTe$_2$, showing that superconductivity is compatible with reduced crystal symmetry \cite{Qi2016,Xie2020,Yang2023}.

We emphasize that the origin of the effect is purely orbital and does not rely on strong spin–orbit coupling. Nevertheless, as illustrated by the comparison between V and Nb, atomic spin-orbit coupling enhances the effect by increasing the magnitude of the induced transverse magnetization.
In this respect, given that strong atomic spin-orbit coupling enhances the proposed effect, the demonstrated synthesis of Ta(111) makes it a promising platform for engineering a robust magnetic response in superconducting systems \cite{Chen2025}.

The net out-of-plane orbital moments for the strained (111) surfaces under an external magnetic field $H=0.1\Delta$ lying within the mirror-plane of the surface can be estimated as
$m^{\mathrm{V}}_{L_c} = -1.47 \times 10^{-5}\,\mu_B$ and
$m^{\mathrm{Nb}}_{L_c} = -2.88 \times 10^{-5}\,\mu_B$.
These resulting magnetic moments are experimentally accessible.
They can be measured using superconducting quantum interference device magnetometry (SQUID), or with momentum resolution by circular dichroism angle-resolved photoemission spectroscopy \cite{Schuler2022}.

Additionally, we propose that the energy-resolved out-of-plane orbital moment can be detected via spin-polarized tunneling currents in scanning tunneling microscopy (STM) or tunneling spectroscopy. Through spin–orbit coupling of the tip, the spin polarization of the tunneling electrons provides access to the orbital-resolved density of states at energies on the order of the superconducting gap, which will manifest through  an asymmetry in the differential conductance $dI/dV$~\cite{Lim2024}.

An interesting consequence of the proposed effect is that the in-plane-field-induced out-of-plane magnetization might offer a novel route to generate superconducting vortices.
Compared with applying a perpendicular field, this mechanism allows the vorticity to be switched on or off by choosing the field orientation relative to the remaining mirror symmetry, enables localized surface vortices since the effect is strongest on the surface, and allows independent tuning of the orbital moment via field strength, direction, and applied strain.
These features may make it a flexible and controllable approach for engineering and manipulating vortices in superconducting devices.

It is worth emphasizing that the effect relies on the multiorbital character of the electronic structure induced by reduced crystalline symmetry, rather than on the presence of multiple bands or Fermi surfaces. While multiband superconductors can support this type of pairing, our results are equally applicable to single-band systems, provided that multiple orbital degrees of freedom are involved.

Finally, we point out that while Berry curvature \cite{Xiao2010} and orbital texture \cite{Mercaldo2023A} may provide a useful framework to relate certain normal--state properties to superconducting response \cite{Gradhand2014-ty}, such as through interband mixing, magnetoelectric couplings \cite{He2021}, or geometric contributions to the superfluid stiffness \cite{Peotta2015}, the emergence of a transverse magnetic response without applied currents is not directly rooted in these effects. 
Instead, our findings reflect a genuinely superconducting phenomenon tied to symmetry breaking and the structure of the condensate itself, which can generate magnetic responses even in the absence of externally driven charge or orbital currents.

\begin{acknowledgment}
M.C., G.C., M.T.M, and C.O. acknowledge partial support from
NRRP MUR project PE0000023-NQSTI. M.C. acknowledges support by Italian Ministry of University and Research (MUR) PRIN 2022 under the Grant No. 2022LP5K7 (BEAT). B.U. acknowledges financial support by the National Research, Development, and Innovation Office (NKFIH Office) of Hungary under Project No. K142652.
\end{acknowledgment}

\section{Methods}

We solve the density-functional Dirac-Bogoliubov-de~Gennes (DBdG) Hamiltonian written in Rydberg units
\begin{equation}
H_{\text{DBdG}}=
 \begin{pmatrix}
   H_D & \Delta_{\text{eff}} \\
   \Delta_{\text{eff}}^\dagger & -H_D^*
 \end{pmatrix},
\end{equation}
where
$
H_D(\mathbf{r}) = 
c \, \boldsymbol{\alpha} \mathbf{p} 
+ \frac{c^2}{2} \left( \boldsymbol{\beta} - \mathbb{I}_4 \right) 
+ \left( V_{\rm eff}(\mathbf{r}) - E_F \right) \mathbb{I}_4 
+ \left( \mathbf{L} \, \mathbb{I}_4 + \boldsymbol{\Sigma} \right) \mathbf{B}_{\rm ext}(\mathbf{r}),
$
with
$\mathbf{\boldsymbol \alpha} = \boldsymbol \sigma_x \otimes \mathbf{\boldsymbol \sigma}$, 
$\boldsymbol \beta = \boldsymbol \sigma_z \otimes \mathbb I_2$, 
$\mathbf{\boldsymbol \Sigma} = \mathbb I_2 \otimes \mathbf{\boldsymbol \sigma}$, 
$\mathbf{\boldsymbol \sigma} $ denotes the Pauli-matrices, 
and $\mathbb I_n $ being the identity matrix of order $n$. $V_{\text{eff}}(\mathbf r)$ and $\mathbf{B}_\text{ext}(\mathbf r)$  are the effective potential and the external field, respectively. $\Delta_\text{eff}(\mathbf r)$ is the effective $4 \times 4$ pairing potential matrix due to the four component Dirac spinors. Importantly, the external magnetic field couples not only to the spin but also to the orbital motion.

The effective electrostatic potential, and pairing potential can be written as
\begin{subequations}
\begin{eqnarray}
  V_{\text{eff}}(\mathbf r)   &=&  
  \int \frac{\rho(\mathbf r')}{|\mathbf r - \mathbf r'|}  \dd^{~\!3}\!r' +
  \frac{\delta E^0_{xc}[\rho]}{\delta \rho(\mathbf r)},\\
  \mathbf \Delta_{\text{eff}}(\mathbf r) &=& \Lambda \, \boldsymbol \chi(\mathbf r),
\end{eqnarray}
\end{subequations}
where $E^0_{xc}[\rho]$ is the exchange correlation functional for the normal state
and $\Lambda$ is the strength of the interaction responsible for superconductivity
(which can be treated as an adjustable site and orbital dependent semi-phenomenological parameter).

We solve the DBdG equations self-consistently with the Screened Korringa-Kohn-Rostoker Green's function (SKKR-GF) method as described for layered systems in Refs.~\onlinecite{Csire2015, Csire2018kkr}.
The central quantity in the KKR method, the Green’s function, is obtained from the generalized multiple scattering theory in a self-consistent manner. 
In this context, the advanced Green’s function acquires additional dimension (Nambu space) and is denoted as:
$
G^{ab,+}_{\alpha \beta}(\varepsilon , \mathbf{r}, \mathbf{r}'),
$
where \(a, b\) refer to the electron-hole (Nambu) space, and \(\alpha, \beta\) represent the spin indices.
Using the Green’s function, the following charge and anomalous densities can be obtained allowing the self-consistent solution:
\begin{equation}
\begin{aligned}
\rho_\alpha(\mathbf r) ={}&
-\frac{1}{\pi} \int_{-\infty}^{\infty} \dd\varepsilon \, f(\varepsilon)\,
\Im \Tr G^{ee,+}_{\alpha\alpha}
(\varepsilon,\mathbf r,\mathbf r) \\[0.4em]
&-\frac{1}{\pi} \int_{-\infty}^{\infty} \dd\varepsilon \,
\bigl(1-f(\varepsilon)\bigr)\,
\Im \Tr G^{hh,+}_{\alpha\alpha}
(\varepsilon,\mathbf r,\mathbf r),
\end{aligned}
\end{equation}
\begin{equation}
\begin{aligned}
&\chi_{\alpha\beta}(\mathbf r) ={}
-\frac{1}{2\pi} \int_{-\infty}^{\infty} \dd\varepsilon\,
\bigl(1-2f(\varepsilon)\bigr) \\[0.4em]
&\times
\Bigl\{
\Im \!\left[
G^{eh,+}_{\alpha\beta}
+ G^{he,+}_{\alpha\beta}
\right] %\\[0.3em]
%&\qquad
+ i \Re \!\left[
G^{he,+}_{\alpha\beta}
- G^{eh,+}_{\alpha\beta}
\right]
\Bigr\}
(\varepsilon,\mathbf r,\mathbf r).
\end{aligned}
\end{equation}

For the effective potentials, we employed the atomic sphere approximation (ASA).
The normal-state electronic structure was calculated self-consistently within the local density approximation (LDA), using the parametrization of Vosko \textit{et al.}\cite{Vosko1980}.
The partial waves within multiple scattering theory are treated with an angular momentum cutoff of $\ell_\mathrm{max}=2$.
In the self-consistent normal state calculations, we used a Brillouin zone (BZ) integration with 200 ${k}$ points in the irreducible wedge of the BZ and a semicircular energy contour on the upper complex plane with 16 points for energy integration. 

After obtaining the self-consistent electrostatic potential in the normal state, superconducting properties were evaluated by solving the KSDBdG equations using the experimental superconducting gap.
We solve the DBdG equations self-consistently, assuming an isotropic $s$-wave spin-singlet BCS superconducting host, with all induced pairing channels incorporated self-consistently.
To ensure convergence, these calculations employed a significantly denser $k$-point mesh of approximately 5000 points around the Fermi surface.
In calculations involving an external magnetic field, the normal-state electrostatic potential was computed without the field, since the applied fields are negligible on the scale of the normal-state electronic structure but relevant on the superconducting energy scale. Accordingly, the magnetic field was included only in the KSDBdG calculations.

\bibliography{main}

\end{document}